\documentclass[acmlarge]{acmart}

\usepackage{booktabs} 

\setcopyright{acmcopyright}

\acmDOI{0000001.0000001}

\received{February 9999}
\received{March 9999}
\received[accepted]{June 9999}

\usepackage{amsmath}
\usepackage{amsfonts}
\usepackage{amssymb}
\usepackage{tikz}
\usetikzlibrary{matrix,positioning,patterns,calc,fit}
\usepackage{pgfplots}
\usepackage{color}
\usepackage{array}
\usepackage{amsthm}
\usepackage{graphicx}
\usepackage{enumerate}
\usepackage{xspace}
\newcommand{\GPSBaseBody}[4]{
    \node (C#4) [fit={(-#2/2, #1/2) (#2/2, -#1/2)}, inner sep=0pt, draw=black, thick] {}; \pgfmatrixnextcell 
    \node {$\mathrel{+}=$}; \pgfmatrixnextcell
    \node (A#4) [fit={(-#3/2, #1/2) (#3/2, -#1/2)}, inner sep=0pt, draw=black, thick] {}; \pgfmatrixnextcell 
    \node (B#4) [fit={(-#2/2, #3/2) (#2/2, -#3/2)}, inner sep=0pt, draw=black, thick] {}; \pgfmatrixnextcell
    \\
}

\newcommand{\GPSBase}[5]{
    \node (#4) [matrix, matrix of nodes, nodes={anchor=center},
                     row sep=0.1cm, column sep=0.1cm,#5] 
                     {\GPSBaseBody{#1}{#2}{#3}{#4}};
}

\newcommand{\GPSPartM}[5]{
    \foreach \a in {1,...,#4}{
        \draw let \p1 = (C#5) in (\x1 - #2/2, \y1 - #1/2 + \a * #1/#4 ) -- ( \x1 + #2/2 , \y1 - #1/2 + \a * #1 / #4); 
    } 
    \foreach \a in {1,...,#4}{
        \draw let \p1 = (A#5) in (\x1 - #3/2, \y1 - #1/2 + \a * #1/#4 ) -- ( \x1 + #3/2 , \y1 - #1/2 + \a * #1 / #4); 
    } 
}

\newcommand{\GPSPartK}[5]{
    \foreach \a in {1,...,#4}{
        \draw let \p1 = (A#5) in (\x1 - #3/2 + \a * #3 / #4, \y1 - #1/2 ) -- ( \x1 - #3/2 + \a * #3 / #4, \y1 + #1/2); 
    } 
    \foreach \a in {1,...,#4}{
        \draw let \p1 = (B#5) in (\x1 - #2/2, \y1 - #3/2 + \a * #3 / #4 ) -- ( \x1 + #2/2, \y1 - #3/2 + \a * #3 / #4 ); 
    }
}

\newcommand{\GPSPartN}[5]{
    \foreach \a in {1,...,#4}{
        \draw let \p1 = (C#5) in (\x1 - #2/2 + \a * #2 / #4, \y1 - #1/2 ) -- ( \x1 - #2/2 + \a * #2 / #4, \y1 + #1/2); 
    } 
    \foreach \a in {1,...,#4}{
        \draw let \p1 = (B#5) in (\x1 - #2/2 + \a * #2 / #4, \y1 - #3/2 ) -- ( \x1 - #2/2 + \a * #2 / #4, \y1 + #3/2); 
    }
}

\newcommand{\GPSShade}[8]{
    \draw let \p1 = (#1) in [#8] (\x1-#3/2+#6*#5, \y1+#2/2-#7*#4) rectangle (\x1-#3/2+#6*#5+#5, \y1+#2/2-#7*#4-#4);
}

\begin{document}
\pgfplotsset{compat=1.12} 
\title{A Tight I/O Lower Bound for Matrix Multiplication}
\author{Tyler Michael Smith}
\affiliation{%
  \institution{ETH Zurich}
  \city{Zurich}
  \state{Zurich}
  \postcode{8006}
  \country{Switzerland}}
\author{Bradley Lowery}
\affiliation{%
  \institution{University of Sioux Falls}
  \city{Sioux Falls}
  \state{SD}
  \postcode{57105}
  \country{USA}}
\author{Julien Langou}
\affiliation{%
  \institution{University of Colorado Denver}
  \city{Denver}
  \state{CO}
  \postcode{80204}
  \country{USA}}
\author{Robert A. van de Geijn}
\affiliation{%
  \institution{The University of Texas at Austin}
  \city{Austin}
  \state{TX}
  \postcode{78712}
  \country{USA}
}

\begin{abstract}
A tight lower bound for required I/O when computing an ordinary matrix-matrix multiplication
on a processor with two layers of memory is established.
Prior work obtained weaker lower bounds by reasoning about the number of \textit{segments} needed to perform $C:=AB$,
for distinct matrices $A$, $B$, and $C$,
where each segment is a series of operations involving $M$ reads and writes
to and from fast memory, and $M$ is the size of fast memory.
A lower bound on the number of segments was then determined by 
obtaining an upper bound on the number of elementary multiplications performed per segment.
This paper follows the same high level approach, but improves the lower bound by 
(1) transforming algorithms for MMM so that they perform all computation via fused multiply-add instructions (FMAs) and
using this to reason about only the cost associated with reading the matrices,
and (2) decoupling the per-segment I/O cost from the size of fast memory.
For $n \times n$ matrices, the lower bound's leading-order term is $2n^3/\sqrt{M}$. 
A theoretical algorithm whose leading terms attains this is introduced.
To what extent the state-of-the-art Goto's Algorithm attains the lower bound is discussed.

\end{abstract}

%
%



\maketitle

\section{Introduction}
Matrix-matrix multiplication (MMM) is an important practical operation from which many applications demand high performance.
A limiting factor on what fraction of theoretical peak this operation can attain is the input-output (I/O) operations
(data movements between memory layers) that are incurred.
For this reason, lower bounds on the I/O requirements are of great interest,
especially as the ratio between the speed of I/O and floating point computation continues to deteriorate.

To derive lower bounds, we start with the assumption that a processor has two layers of memory hierarchy,
a small {\em fast} memory and large {\em slow} memory.
The fast and slow memory could represent the cache(s) and main memory of a processor,
respectively, or main memory and disk.
Practical implementations minimize the movement of data between these (and more) layers.  

In this paper, we only consider conventional or ordinary matrix-matrix multiplication where $ C := A B $,
requiring $mnk$ elementary multiplications and $mn\left(k-1\right)$ elementary additions.
The quantities $m$, $n$, and $k$ are always used to refer to the sizes of $C$, $A$, and $B$,  
which are $ m \times n $, $ m \times k $, and $ k \times n $, respectively.
The quantity $M$ always refers to the capacity of fast memory, in elements.
We assume that $A$, $B$, and $C$ are distinct matrices.

This paper advances upon the state-of-the-art to obtain the exact coefficient for the leading term of I/O lower bound.
The improvement over prior lower bounds comes mainly from two contributions. 
\begin{enumerate}
\item It shows that any algorithm for MMM can be transformed into one that casts all computation in terms of FMA instructions.
This gives the same benefit as in~\cite{dongarra2008masterworker},
but without requiring that absolutely all computation be performed via FMA instructions.
\item Prior works~\cite{redblue,irony2004communication,dongarra2008masterworker} have obtained lower bounds on MMM by
breaking computation into segments of I/O cost equal to the capacity of fast memory.
This present paper shows that I/O lower bounds can be improved by turning this I/O cost per segment into a free variable.
\end{enumerate}
This paper proceeds to discuses algorithms that attain 
the lower bound when ignoring lower ordered terms, showing that the leading term on the lower bound is sharp.
It shows that the practical algorithm in Goto and van de Geijn~\cite{GotoBLAS} gets close to the lower bound
for some levels of cache but not others.
Together, this advances the understanding of the limits on performance for this operation.

\section{Problem Definition and Model of Computation}
In this section, we give a formal definition and model of computation for MMM for which we will derive I/O lower bounds.
We write $a_{ij}$ or $b_{ij}$ to denote the element in the $i^{th}$ row and $j^{th}$ column of the matrix $A$ or $B$ respectively,
and we write $c_{ij}$ to denote a nontrivial partial sum of the element in the $i^{th}$ row and $j^{th}$ column of the matrix $C$.

\subsection{Problem Definition} 

We first begin by defining the operations of interest and the elementary instructions that are used to create algorithms to implement them.
In order to balance the number of elementary multiplication and elementary additions,
work with the matrix-matrix multiplication and accumulation (MMMA) operation instead.
\begin{definition}[MMMA]
An MMMA operation with problem size $m, n, k$ computes $C \mathrel{+}= AB$.
with each elementary multiplication $c_{ijp} = a_{ip} b_{p j}$ performed explicitly,
and $c_{ij}$ is formed by summing over $c_{ijp}$ for all $p$, plus the initial $c_{ij}$.
There are exactly $mnk$ distinct elementary multiplications in an MMMA operation.
\label{def:mmma}
\end{definition}
This only defines what elementary computations must be performed during an MMMA operation,
and does not impose any requirement on how they are computed.
Most notably, this definition allows the creation of intermediate quantities that are later summed to form an element of $C$.

\subsection{Model of Computation}
We now introduce five types of instructions that operate on variables.
A variable is an element of $A$, $B$ or $C$, or partial sum of an element of $C$, i.e. a summation over $c_{ijp}$ for some but not all $p$.
\begin{enumerate}
\item \textbf{Read.} Move variable from slow memory into fast memory.
\item \textbf{Write.} Move a variable from fast memory to slow memory.
\item \textbf{Multiply.} If $a_{ip}$ and $b_{p j}$ are in fast memory, multiply them.
    This creates a new variable $c_{ijp}$ in fast memory. 
    There must be space in fast memory to accommodate $c_{ijp}$.
\item \textbf{Add.} If $c_1$ and $c_2$ are in fast memory, add them, storing the result in the variable $c_1$.
\item \textbf{FMA.} If variables $a_{ip}$, $b_{p j}$ and $c_{ij}$ are in fast memory,
      multiply $a_{ip}$ and $b_{p j}$ and add them to $c_{ij}$, without requiring any additional space in fast memory.
\item \textbf{Delete.} Delete a variable so that it is no longer in fast memory. 
\end{enumerate}
With this, an algorithm for MMMA is a sequence of the above types of instructions that performs an MMMA operation.
The read cost of an algorithm is the total number of read instructions.
The write cost is the total number of write instructions,
and the I/O cost of an algorithm is the sum of its read and write costs.

In our model of computation, we allow caches to be warm, i.e. 
at the beginning and at the end of computation, fast memory can contain whatever elements of $A$, $B$, and $C$.

We assume that any algorithm for MMMA does not perform any computation more than once,
but without any further restriction on generality.
The only computation modeled by our MMMA algorithms are addition and multiplication,
and the result of each multiplication and addition only contributes to a single element of the output.
If any value is ever computed twice, one of the results can be discarded.

\section{Finding the I/O Lower Bound}
We now use our definition and model of computation to obtain a lower bound on the number of reads that any algorithm for MMMA must perform.
The fact that Definition~\ref{def:mmma} allows the creation of intermediate quantities makes it difficult to reason about how many times
elements of $C$ must be read from fast memory.
To address this difficulty we first transform algorithms for MMMA
such that all computation is performed via FMAs, taking an element each of $A$, $B$, and $C$ as inputs.
We then place a lower bound on the read cost on the transformed algorithms.

\begin{lemma}
\label{lemma:fma_transformation}
Given any algorithm for MMMA, there exists a valid algorithm for MMMA
where all computation is performed via $mnk$ FMA instructions of the form $c_{ij} \mathrel{+}= a_{ip} b_{pj}$,
no variables are created that store intermediate quantities of elements of $C$,
and the number of read instructions in the second algorithm is at most the number of read instructions in the first.
\end{lemma}
\begin{proof}
Consider an algorithm for MMMA where there exists a multiplication instruction whose output is not immediately added to $c_{ij}$,
where $c_{ij}$ is the variable that will ultimately accumulate the final result of an element of $C$.
If there are any temporary variables representing partial sums of that same element,
one of them is directly added to $c_{ij}$.
Then the algorithm contains a subsequence of instructions starting with the multiplication
$c_{ijp} \mathrel{:}= a_{ip} b_{p j}$ and ending with $c_{ij} \mathrel{+}= c_{ijp}$.
The following a transformation eliminates the temporary variable $c_{ijp}$.
\begin{enumerate}
\item If in the original algorithm $c_{ij}$ is not in fast memory when the multiplication instruction occurs,
insert a read instruction to move $c_{ij}$ into fast memory immediately before the multiplication instruction.
This increases the number of read instructions by one if and only if $c_{ij}$ was not already in fast memory.
This read instruction requires space in fast memory for a single element,
but there is such a space because the multiplication instruction in the original algorithm requires it.
\item Replace $c_{ijp} \mathrel{:}= a_{ip} b_{p j}$ with $c_{ij} \mathrel{+}= a_{ip} b_{p j}$.
\item When the addition occurs in the original algorithm, both $c_{ijp}$ and $c_{ij}$ must be in fast memory.
Because of this, if $c_{ij}$ was not in fast memory when the multiplication occurred, there exists at least one read instruction that
loads it into fast memory between the multiplication instruction and the addition instruction.
One of these is redundant, so delete the latest instruction that read either $c_{ijp}$ or $c_{ij}$ into fast memory 
between the multiplication instruction and the addition instruction.
\item Remove the addition instruction $c_{ij} \mathrel{+}= a_{ip} b_{pj}$.
\item Remove any delete instructions referencing $c_{ijp}$ as the variable never exists in the transformed algorithm.
\item Replace any occurrences of $c_{ijp}$ in any computation instructions in the algorithm with $c_{ij}$.
\item Finally we modify any read and write instructions to ensure that $c_{ij}$ is in fast memory in the transformed algorithm
whenever $c_{ijp}$ is in fast memory in the original algorithm.
We inspect the original algorithm for contiguous subsequences during which both $c_{ij}$ and $c_{ijp}$ are both in fast memory 
(ignoring the subsequence beginning with the multiplication instruction that creates $c_{ijp}$, which we have already handled).
Each begins with a read instruction reading the second of the two variables and ends with a write instruction writing the first of the two.
Delete those two instructions, and replace all other read and write instructions that read and write $c_{ijp}$ with instructions that read and write $c_{ij}$.
\end{enumerate}

The transformation may add a single read instruction, but if it does so, the transformation deletes one other read instruction
and so the total number read instructions is unchanged. 
The transformation does not affect the number of writes.
At no point does the transformation increase the footprint of the algorithm on fast memory.
Finally, the algorithm still correctly performs an MMMA operation. Whenever $c_{ijp}$ is used as an accumulator,
$c_{ij}$ is used instead, and $c_{ij}$ is in fast memory in the transformed algorithm whenever $c_{ijp}$ is in fast memory in the original algorithm,
so it is always valid to reference it.
After the transformation is applied, it can be successively applied to every multiplication so that all computation is performed via FMA instructions.
\end{proof}

We first break the computation into \textit{segments}, where we know the number of read instructions in each segment.
Then a lower bound on the number of reads can be found via a lower bound on the number of segments.
\begin{definition}[Segment]
Divide an MMMA algorithm into contiguous subsequences 
such that the subsequences are adjoining,
and each subsequence but perhaps the last one has exactly $R$ read instructions.
Then the subsequences are called segments of read cost $R$.
\label{def:segment}
\end{definition}

\begin{lemma}
Let $F_{M+R}$ be an upper bound on the number of distinct FMA instructions executed during any segment of read cost $R$ in any MMMA algorithm.
Then any MMMA algorithm must have a total read cost of at least $$R\left(\frac{mnk}{F_{M+R}}-1\right).$$
\label{lemma:lower_bound_sin_f}
\end{lemma}
\begin{proof}
By Lemma~\ref{lemma:fma_transformation} any MMMA algorithm can be transformed into an algorithm
with the same read cost that casts all of its computation in terms of FMA instructions.
This is true even if the original is executed on a machine without access to FMAs.
Therefore a lower bound on the transformed algorithm is a lower bound on the original one.
The transformed algorithm has exactly $mnk$ FMA instructions so there are at least $\frac{mnk}{F_{M+R}}$ segments.
All but the last segment have a read cost of exactly $R$, so the transformed algorithm and hence the original algorithm
have a read cost of at least $R \left(\frac{mnk}{F_{M+R}}-1\right)$.
\end{proof}

Together, Definition~\ref{def:segment} and Lemma~\ref{lemma:lower_bound_sin_f}
represent a simplification of the S-partitioning problem, introduced in~\cite{redblue}
and subsequently used in other I/O complexity lower bounds for MMM~\cite{irony2004communication,dongarra2008masterworker}.
Segments are very similar to the \textit{subcalculations} from the S-Span theorem in~\cite{redblue}
and \textit{phases} from~\cite{irony2004communication}.
However, our segments are defined solely by the number of reads,
whereas the others are defined by the number of reads plus writes.
Additionally, while each segment (except the last) has the same number of read instructions,
the fact that that number is a free variable distinguishes our segments from prior work.

An upper bound on the amount of computation per segment, $F_{M+R}$, can be obtained by considering
the number of elements of $A$, $B$, and $C$ that a segment has access to.
There are a total of $M$ elements in fast memory when a segment begins and there are a total of $R$ elements read into fast memory during a segment.
The following geometric inequality by Loomis and Whitney~\cite{loomiswhitney} can be used to 
place an upper bound on the number of FMAs that can be performed using $M+R$ elements as inputs.

\begin{theorem}[Discrete Loomis-Whitney Inequality]
Let $V$ be a finite set with elements in $\mathbb Z^3$, and let $V_x$, $V_y$, and $V_z$ be orthogonal projections of $V$ onto the coordinate planes.
Then the cardinality of $V$, $\left| V \right|$, satisfies
$$\left|V\right| \leq \sqrt{ \left| V_x \right| \left| V_y \right| \left| V_z \right|}$$
\label{theorem:loomis_whitney}
\end{theorem}

\begin{lemma}
    Using $N_A$ distinct elements of $A$, $N_B$ distinct elements of $B$, and $N_C$ elements of $C$ as inputs,
    one can perform at most $\sqrt{N_A N_B N_C}$ distinct FMAs.
    \label{lemma:applyLW}
\end{lemma}
\begin{proof}
Suppose there is a set of FMAs that can be performed using $N_A$, $N_B$, and $N_C$ elements of $A$, $B$, and $C$, respectively.
Then the set of FMAs can be identified with a set, $V$, in $\mathbb{Z}^3$, where each FMA $c_{xy} \mathrel{+}= a_{xz} \cdot b_{zy}$ has the coordinate $(x, y, z)$.
In order to execute an FMA corresponding to the coordinate $(x,y,z)$,
we must use the element $c_{xy}$, identified with the coordinate $(x,y)$,
the element $a_{xz}$ with coordinate $(x,z)$, and the element $b_{zy}$ with coordinate $(z,y)$.

By projecting $V$ in the $z$ dimension, we obtain a set of points $(x,y)$ in $\mathbb{Z}^2$, 
corresponding to the set of elements of $C$ that are used in order to execute the FMAs.
Similarly, if we project $V$ along the other coordinate axes, we obtain sets of elements of $A$ and $B$ that are used in order to execute an FMA.
These three orthogonal projections have cardinalities of at most $N_A$, $N_B$, and $N_C$.
By Lemma~\ref{theorem:loomis_whitney} $|V|$, the number of FMAs one can perform, is at most $\sqrt{N_A N_B N_C}$.
\end{proof}
The use of the Loomis-Whitney inequality for MMM I/O lower bounds,
Lemma~\ref{lemma:applyLW} and its proof first appeared in Irony et al.~\cite{irony2004communication}.

\begin{lemma}
$F_{M+R}$, the maximum number of distinct FMA instructions executed during a segment of read cost $R$ in any MMMA algorithm,
is at most $\left(\frac{1}{3}(M+R)\right)^{3/2}$.
\label{lemma:F_upper_bound}
\end{lemma}
\begin{proof}
In order to find an upper bound on the number of FMAs that can be executed during a segment,
we will use two constraints: One from the size of fast memory, and one from the number of loads.
\begin{enumerate}
    \item Let $M_A$, $M_B$, and $M_C$ be, respectively, the number of elements of $A$, $B$, and $C$ in fast memory
    at the start of a segment. This gives us the constraint from the size of fast memory $M_A + M_B + M_C \leq M$.
    \item Let $R_A$, $R_B$, and $R_C$ be, respectively, the number of elements of $A$, $B$, and $C$ loaded from slow memory
    during a segment. This gives us the constraint from the number of loads $R_A + R_B + R_C \leq R$.
\end{enumerate}
Let $N_A = M_A + R_A$, $N_B = M_B + R_B$, and $N_C = M_C + R_C$.
Adding constraints (1) and (2) gives us $N_A + N_B + N_C \leq M+R$.
We can find an attainable upper bound on $F_{M+R}$ by combining this constraints with Lemma~\ref{lemma:applyLW},
giving us the following problem:
\[
\mbox{Maximize }F_{M+R}
\mbox{ under the constraints }
\left\{ 
\begin{array}{r@{~}c@{~}l}
F_{M+R} &\leq& \sqrt{N_A N_B N_C}\\
0 &\leq& N_A, N_B, N_C \\
N_A + N_B + N_C &\leq& M+R.
\end{array} \right. 
\]
Application of the Lagrange multiplier method, detailed in Appendix~\ref{maximum_xyz}, tells us that the 
global maximum occurs when 
\[
N_A = N_B = N_C = \frac{1}{3}(M+R)
\quad \mbox{so that} \quad
F_{M+R} \le \left(\frac{1}{3}(M+R)\right)^{3/2}.
\]
\end{proof}

From Lemma~\ref{lemma:lower_bound_sin_f} and~\ref{lemma:F_upper_bound},
we know that any algorithm for MMMA has a read cost of at least
$\left( \frac{mnk}{\left(\frac{1}{3}(M+R)\right)^{3/2}} - 1 \right) R.$
Lemma~\ref{lemma:F_upper_bound} for the case where $R$ always equal to $M$ can be found in~\cite{dongarra2008masterworker},
yielding an I/O lower bound of $\frac{3 \sqrt{3}}{ 2 \sqrt{2} } \frac{mnk}{\sqrt{M}} - M$.

Our lower bound is dependent on $R$, which is a free variable.
In order to find the greatest lower bound for large problem sizes,
we want the positive $R$ that maximizes
$\left( \frac{R mnk }{\left(\frac{1}{3}(M+R)\right)^{3/2}}\right)$.
This occurs when $R = 2M$, yielding the following theorem.

\begin{theorem}
Any algorithm for MMMA on a machine with fast memory of capacity $M$ has a read cost of at least
$$ \frac{2mnk}{\sqrt{M}} - 2M. $$
\end{theorem}

If we define the MMM operation, $C \mathrel{:}= AB$ the same way as MMMA,
except that it does not need to add the initial $c_{ij}$, we obtain the following corollary.
\begin{corollary} Any algorithm for MMM on a machine with fast memory of capacity $M$ has a read cost of at least
$$ \frac{2mnk}{\sqrt{M}} - mn - 2M. $$
\end{corollary}
\begin{proof}
Given any algorithm for MMM, there exists an algorithm for MMMA that has exactly $mn$ additional read instructions.
Consider the multiply instruction in the original MMM algorithm
that creates the variable that will ultimately contain the element in the $i^{th}$ row and $j^{th}$ column of $C$.
We can replace this instruction with a load instruction that loads the initial element in the
$i^{th}$ row and $j^{th}$ of $C$, followed by an FMA instruction.

Doing this for each element of $C$ yields an algorithm for MMMA.
This algorithm has a read cost of at least $\frac{2mnk}{\sqrt{M}} - 2M$,
so the algorithm for MMM has a read cost of at least $\frac{2mnk}{\sqrt{M}} - mn - 2M$,
\end{proof}

Because we argued exclusively about the inputs to computation,
we obtained a lower bound on the read cost by itself.
Therefore we can add a separate lower bound on the number of writes to obtain an I/O lower bound,
and any algorithm for MMMA has at least $mn - M$ compulsory writes 
(or $mn$ if all elements of $C$ must be in slow memory at the end of execution).
\begin{corollary}
Any algorithm for MMMA on a machine with fast memory of capacity $M$ has an I/O cost of at least
$$ \frac{2mnk}{\sqrt{M}} + mn - 3M. $$
Any algorithm for MMM on a machine with fast memory of capacity $M$ has an I/O cost of at least
$$ \frac{2mnk}{\sqrt{M}} - 3M. $$
\end{corollary}

\section{Attaining the I/O Lower Bound}
In this section, we first develop intuition for how to attain near-optimal I/O cost for MMM algorithms.
This motivates an algorithm that attains the lower bound for a processor with two layers of memory, a fast and a slow memory.
Finally, we discuss the state-of-the-art algorithm by Goto~\cite{GotoBLAS} in the context of our lower bound.

\subsection{Blocked algorithms}

It has been well-known since the arrival of hierarchical memories in the 1980s
that high-performance implementations of many dense linear algebra algorithms,
including MMM, are facilitated by so-called blocked algorithms.
The reason is simple: Multiplying $ C := A B + C $ when $ m = n = k = n_b $ and all matrices fit into fast memory 
allows $ \mathcal{O}( n_b^2 ) $ reads and writes to be amortized over $ 2 n_b^3 $ scalar floating point operations (flops).
When $ m $, $ n $, $ k $ are larger than $ n_b $, the operands can be blocked into $ n_b \times n_b $ submatrices and staged as a sequence of 
MMMA operations with these submatrices.
The question is how to optimally break the operands into submatrices.

\subsection{Insights}
\label{optimal_properties}

From the I/O lower bound $Q_{AB + C} \geq {2mnk}/{\sqrt{M}} + mn - 3M$,
and from the fact that an MMMA requires $mnk$ FMAs,
we know that if an optimal algorithm exists, it must perform $\sqrt{M}/{2}$ FMAs per read and write for large matrices.
Rewriting the ratio as $M$ FMAs per $2\sqrt{M}$ reads and writes, the following points towards an algorithm:
Assuming a $\sqrt{M} \times \sqrt{M}$ block of $C$ is already in fast memory,
perform a rank-1 update of that block of $C$ memory using
the appropriate part of a column of $ A $ with $\sqrt{M}$ elements and the appropriate part of a row of $ B $ with
$\sqrt{M}$ elements.  This requires $ 2 \sqrt{M} $ reads and writes for $M$ FMAs.

Fast memory cannot hold  all of these elements at once.
However it is possible to do so with a $(\sqrt{M}-1) \times (\sqrt{M}-1)$ block of $C$ instead,
leaving enough room for $ \sqrt{M}-1 $ elements of $ A $ and $ \sqrt{M}-1 $ elements of $ B $, 
thus achieving close to the desired I/O lower bound.
By performing the update of the block of $ C $ as a sequence of rank-1 updates,
bringing that block of $ C $ into fast memory can be amortized over many flops,
thus essentially achieving the assumption that the block is already in fast memory.

\subsection{An asymptotically optimal blocked algorithm}
\begin{figure}
\begin{center}
\newcommand{\partx}[4]{
    \foreach \a in {1,...,#2}{
        \draw let \p1 = (#1) in (\x1 + \a * #4, \y1 ) -- ( \x1 + \a * #4, \y1 - #3); 
    }   
}
\newcommand{\party}[4]{
    \foreach \a in {1,...,#2}{
        \draw let \p1 = (#1) in (\x1, \y1 - \a * #3 ) -- ( \x1 + #4, \y1 - \a * #3); 
    }   
}
\begin{tikzpicture}
\node (RCLABEL) {Algorithm C};
\node [below=3.00cm of RCLABEL] (RBLABEL) {Algorithm B};
\node [below=3.00cm of RBLABEL] (RALABEL) {Algorithm A};

\newcommand{\Cm}{2.75cm}
\newcommand{\Cn}{2.75cm}
\newcommand{\Ck}{3cm}
\newcommand{\nparts}{4}
\GPSBase{\Cm}{\Cn}{\Ck}{Cres}{right=.7cm of RCLABEL}
\GPSPartM{\Cm}{\Cn}{\Ck}{\nparts}{Cres}{}
\GPSPartN{\Cm}{\Cn}{\Ck}{\nparts}{Cres}{}
\GPSShade{CCres}{\Cm}{\Cn}{\Cm/\nparts}{\Cn/\nparts}{1}{1}{fill=magenta}
\path (ACres) +(-\Ck/2, \Cm/2 - \Cm / \nparts) coordinate (ACresPartk);
\partx{ACresPartk}{20}{\Cm / \nparts}{\Ck / 20}
\path (BCres) +(-\Cn/2 + \Cn / \nparts, \Ck/2) coordinate (BCresPartk);
\party{BCresPartk}{20}{\Ck / 20}{\Cm / \nparts}
\draw let \p1 = (ACresPartk) in [fill=magenta] (\x1+4*\Ck/20,\y1) rectangle (\x1+5*\Ck/20,\y1-\Cm/\nparts);
\draw let \p1 = (BCresPartk) in [fill=magenta] (\x1,\y1-4*\Ck/20) rectangle (\x1+\Cn/\nparts,\y1-5*\Ck/20);

\newcommand{\Bm}{3cm}
\newcommand{\Bn}{2.75cm}
\newcommand{\Bk}{2.75cm}
\GPSBase{\Bm}{\Bn}{\Bk}{Bres}{right=.7cm of RBLABEL}
\GPSPartK{\Bm}{\Bn}{\Bk}{\nparts}{Bres}{}
\GPSPartN{\Bm}{\Bn}{\Bk}{\nparts}{Bres}{}
\GPSShade{BBres}{\Bk}{\Bn}{\Bk / \nparts}{\Bn / \nparts}{1}{1}{fill=magenta}
\path (CBres) +(-\Bn/2 + \Bn / \nparts, \Bm/2) coordinate (CBresPartm);
\party{CBresPartm}{20}{\Bm / 20}{\Bn / \nparts}
\path (ABres) +(-\Bk/2 + \Bk / \nparts, \Bm/2) coordinate (ABresPartm);
\party{ABresPartm}{20}{\Bm / 20}{\Bk / \nparts}
\draw let \p1 = (ABresPartm) in [fill=magenta] (\x1,\y1-4*\Bm/20) rectangle (\x1+\Bk/\nparts,\y1-5*\Bm/20);
\draw let \p1 = (CBresPartm) in [fill=magenta] (\x1,\y1-4*\Bm/20) rectangle (\x1+\Bn/\nparts,\y1-5*\Bm/20);

\newcommand{\Am}{2.75cm}
\newcommand{\An}{3cm}
\newcommand{\Ak}{2.75cm}
\GPSBase{\Am}{\An}{\Ak}{Ares}{right=.7cm of RALABEL}
\GPSPartM{\Am}{\An}{\Ak}{\nparts}{Ares}{}
\GPSPartK{\Am}{\An}{\Ak}{\nparts}{Ares}{}
\GPSShade{AAres}{\Am}{\Ak}{\Am / \nparts}{\Ak / \nparts}{1}{1}{fill=magenta}
\path (CAres) +(-\An/2, \Am/2 - \Am / \nparts) coordinate (CAresPartn);
\partx{CAresPartn}{20}{\Am / \nparts}{\An / 20}
\path (BAres) +(-\An/2, \Ak/2 - \Ak / \nparts) coordinate (BAresPartn);
\partx{BAresPartn}{20}{\Ak / \nparts}{\An / 20}
\draw let \p1 = (CAresPartn) in [fill=magenta] (\x1+4*\An/20,\y1) rectangle (\x1+5*\An/20,\y1-\Am/\nparts);
\draw let \p1 = (BAresPartn) in [fill=magenta] (\x1+4*\An/20,\y1) rectangle (\x1+5*\An/20,\y1-\Ak/\nparts);

\node (cache) [fill=magenta, fit={(-0.4cm/2, -0.4cm/2) (0.4cm/2, 0.4cm/2)}, inner sep=0pt, draw=black, thick, yshift=-.75cm] at (Ares.south) {};
\node (mem) [fill=none,  fit={(-0.4cm/2, -0.4cm/2) (0.4cm/2, 0.4cm/2)}, inner sep=0pt, draw=black, thick, yshift=-.3cm] at (cache.south) {};
\node [right] at (cache.east) { {\scriptsize Data in cache.} };  
\node [right] at (mem.east) { {\scriptsize Data in main memory.} };  
\end{tikzpicture}
\end{center}
\caption{Three algorithms for matrix multiplication that attain the lower bound for a single level of cache.}

\label{fig:ABC}
\end{figure}
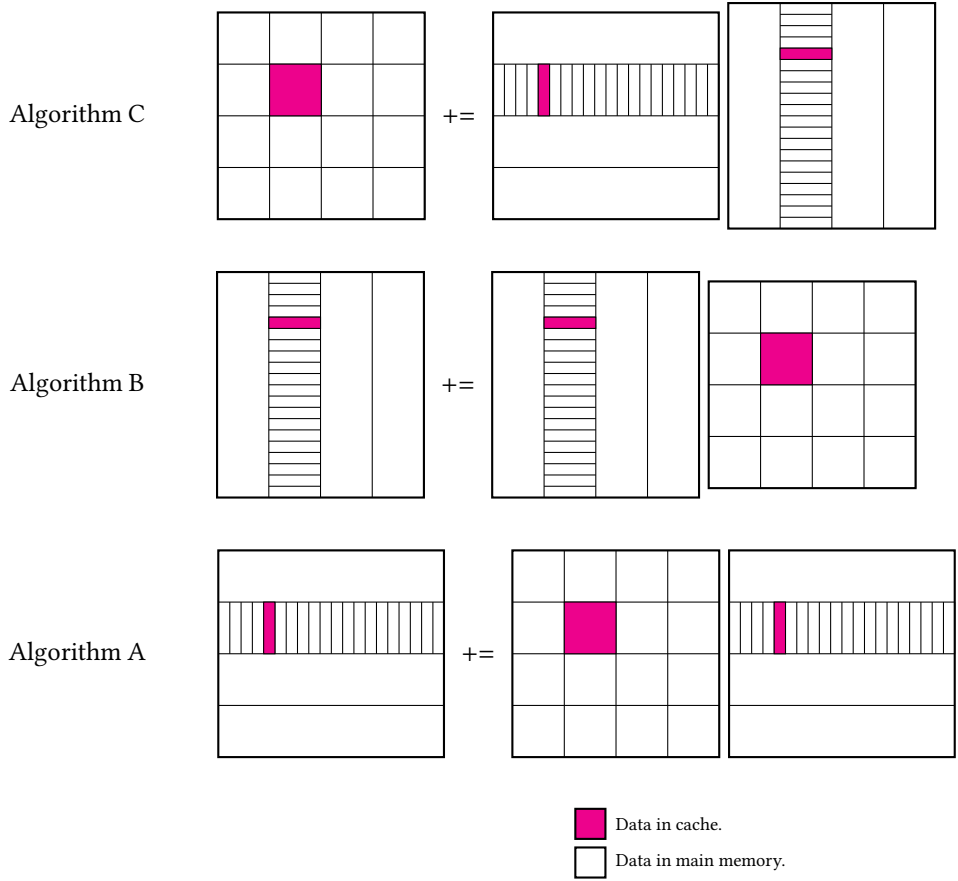

\paragraph*{Algorithm C} 
We are now ready to give an optimal  algorithm (in the sense of asymptotically attaining the lower bound).  Consider $ C := A B + C $.  Partition:
\[
C \rightarrow 
\left( \begin{array}{c | c | c}
C_{0,0} & \cdots & C_{0,N-1} \\ \hline
\vdots & & \vdots \\  \hline
C_{M-1,0} & \cdots & C_{M-1,N-1} 
\end{array}
\right), 
\quad
A \rightarrow
\left( \begin{array}{c}
A_0 \\ \hline
\vdots \\ \hline
A_{M-1}
\end{array}
\right),
\quad 
B \rightarrow 
\left( \begin{array}{c | c | c}
B_0 & \cdots & B_{N-1} 
\end{array}
\right),
\]
where $ C_{i,j} $ is $ (\sqrt{M}-1) \times (\sqrt{M}-1) $,
$ A_i $ is $ (\sqrt{M}-1) \times k$, and $ B_j $ is $ k \times (\sqrt{M}-1) $.
Then a simple loop over all the blocks of $ C $, computing $ C_{i,j} = A_i B_j + C_{i,j} $ via rank-1 updates, requires
\begin{eqnarray*}
\frac{m}{(\sqrt{M}-1)}
\frac{n}{(\sqrt{M}-1)} \left(
2 (\sqrt{M}-1)k + (\sqrt{M}-1)^2 \right) 
=
2 \frac{m n k}{\sqrt{M}-1}
 + mn
\approx
2 \frac{m n k}{\sqrt{M}}
 + mn
\end{eqnarray*}
reads and $ m n $ writes.
When $ mnk $ is large, this algorithm (illustrated in Figure~{\ref{fig:ABC}})  approaches the lower bound.

\subsection{Read-optimal and write-hidden algorithms}

\paragraph*{Algorithm B}
An early occurrence of an algorithm that we can now realize is optimal in terms of the number of reads was presented and analyzed in~\cite{lam1991cache}.
Partition:
\[
C \rightarrow 
\left( \begin{array}{c | c | c}
C_0 & \cdots & C_{N-1} 
\end{array}
\right),
\quad
A \rightarrow 
\left( \begin{array}{c | c | c}
A_0 & \cdots & A_{N-1} 
\end{array}
\right),
\quad
B \rightarrow 
\left( \begin{array}{c | c | c}
B_{0,0} & \cdots & B_{0,N-1} \\ \hline
\vdots & & \vdots \\  \hline
B_{M-1,0} & \cdots & B_{M-1,N-1} 
\end{array}
\right),
\]
where $C_j$ and $ A_p $ are $ m \times  (\sqrt{M}-1) $, and
$ B_{p,j} $ is $ (\sqrt{M}-1) \times  (\sqrt{M}-1) $.
Then a simple loop over all blocks of $ B $,
keeping $ B_{p,j} $ in fast memory and streaming rows of $ C_j $ and $ A_p $ while
computing $ C_j := A_p B_{p,j} + C_j $, requires
\[
\frac{k}{(\sqrt{M}-1)}
\frac{n}{(\sqrt{M}-1)} \left(
2 m (\sqrt{M}-1) + (\sqrt{M}-1)^2 \right) 
\approx 
2 \frac{m n k}{\sqrt{M}} + nk
\]
reads and
\[
\frac{k}{(\sqrt{M}-1)}
\frac{n}{(\sqrt{M}-1)} 
m (\sqrt{M}-1) 
\approx 
\frac{m n k}{\sqrt{M}} 
\]
writes.

The read cost of this algorithm, illustrated in Figure~\ref{fig:ABC}, is essentially equal to the I/O lower bound,
but it requires many writes to slow memory and so cannot be considered I/O optimal.
On the other hand, processors often have full-duplex memory bandwidth 
(meaning that the bandwidth available for reads is separate from the bandwidth available for writes),
so the write cost may not be visible if it is less than or equal to than the read cost and if the reads and writes can be overlapped.
Since that is the case for this algorithm,
it may execute just as efficiently as the algorithm presented in Section~\ref{optimal_properties}.
Thus, we can say that this algorithm is read-optimal and write-hidden.
This becomes important when we later discuss practical implementations.

\paragraph*{Algorithm A} We now present an algorithm that is in some sense the mirror image to Algorithm B,
keeping a square block of $A$ in fast memory instead instead of a square block of $B$.
Partition:
\[
C \rightarrow 
\left( \begin{array}{c}
C_0 \\ \hline
\vdots \\ \hline
C_{M-1} 
\end{array}
\right),
\quad
A \rightarrow 
\left( \begin{array}{c | c | c}
A_{0,0} & \cdots & A_{0,K-1} \\ \hline
\vdots & & \vdots \\  \hline
A_{M-1,0} & \cdots & A_{K-1,N-1} 
\end{array}
\right),
\quad
B \rightarrow 
\left( \begin{array}{c}
B_0 \\ \hline
\vdots \\ \hline
B_{K-1} 
\end{array}
\right),
\]
where $C_i$ and $ B_p $ are $   (\sqrt{M}-1) \times n $, and
$ A_{i,p} $ is $ (\sqrt{M}-1) \times  (\sqrt{M}-1) $.
Then a simple loop over all blocks of $ A $,
keeping $ A_{i,p} $ in fast memory and streaming columns of $ C_i $ and $ B_p $ while
computing $ C_i := A_{i,p} B_{p} + C_i $ yields an algorithm with I/O cost of
\[
\frac{k}{(\sqrt{M}-1)}
\frac{n}{(\sqrt{M}-1)} \left(
2 m (\sqrt{M}-1) + (\sqrt{M}-1)^2 \right) 
\approx 
2 \frac{m n k}{\sqrt{M}} + mk
\]
reads and
\[
\frac{k}{(\sqrt{M}-1)}
\frac{n}{(\sqrt{M}-1)} 
m (\sqrt{M}-1) 
\approx 
\frac{m n k}{\sqrt{M}} 
\]
writes.
The algorithm is also illustrated in Figure~\ref{fig:ABC}.

\subsection{Practical implications}

We now discuss how current practical algorithms measure up against the theoretical lower bound.

\subsubsection{Goto's Algorithm}

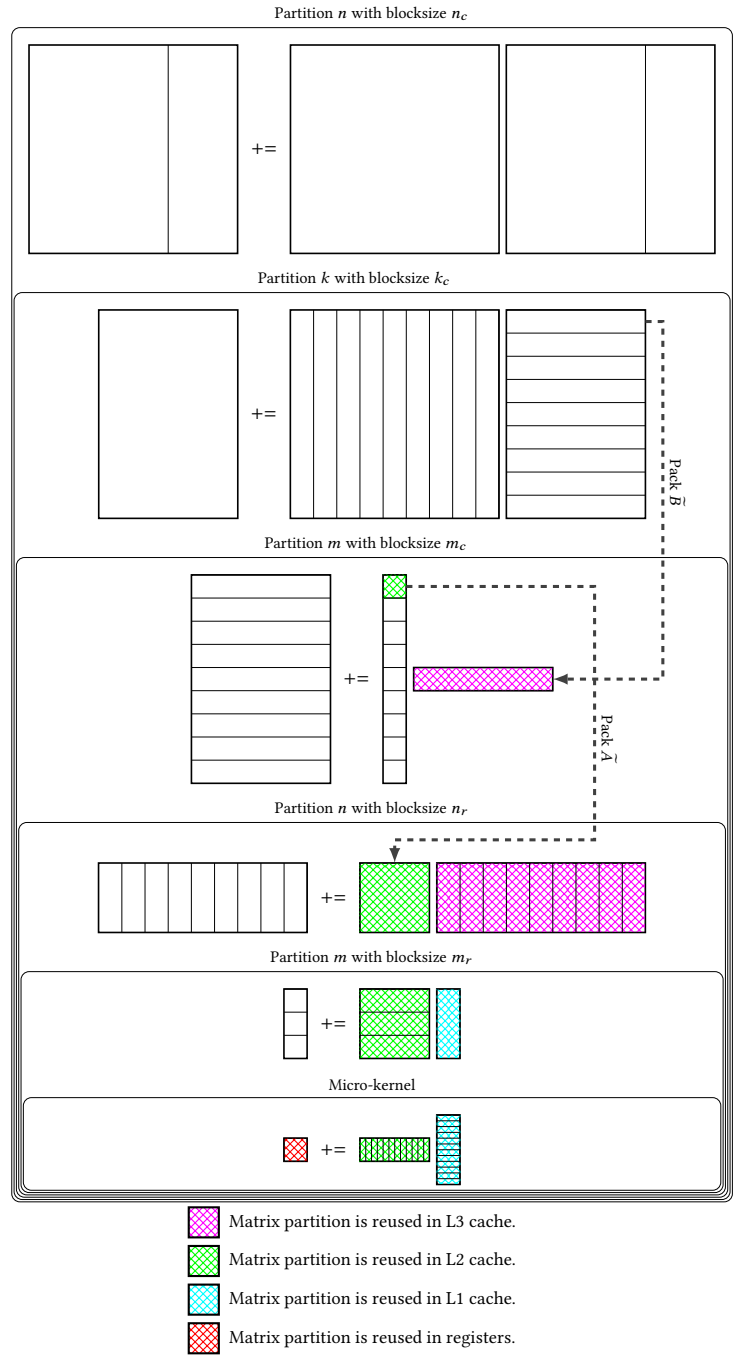
\begin{figure}
\centering
\resizebox{!}{16cm}{
\begin{tikzpicture}[>=latex,node distance=-0.25cm and .00cm]
\def\dbig{3.6cm}
\def\dmed{1.2cm}
\def\dsma{0.4cm}
\GPSBase{\dbig}{\dbig}{\dbig}{l3n}{}
\draw let \p1 = (Cl3n) in (\x1 - \dbig/2+2*\dbig/3,\y1-\dbig/2) -- (\x1-\dbig/2+2*\dbig/3,\y1+\dbig/2);
\draw let \p1 = (Bl3n) in (\x1 - \dbig/2+2*\dbig/3,\y1-\dbig/2) -- (\x1-\dbig/2+2*\dbig/3,\y1+\dbig/2);
\GPSBase{\dbig}{2/3 * \dbig}{\dbig}{l3k}{below=0.7cm of l3n}
\GPSPartK{\dbig}{2/3 * \dbig}{\dbig}{9}{l3k}
%
\GPSBase{\dbig}{2/3 * \dbig}{\dsma}{l2m}{below=0.7cm of l3k}
\GPSShade{Bl2m}{\dsma}{2/3 * \dbig}{\dsma}{2/3 * \dbig}{0}{0}{pattern=crosshatch, pattern color=magenta}
\GPSPartM{\dbig}{2/3*\dbig}{\dsma}{9}{l2m}
\GPSShade{Al2m}{\dbig}{\dsma}{\dsma}{\dsma}{0}{0}{pattern=crosshatch, pattern color=green}
\GPSBase{\dmed}{\dbig}{\dmed}{innerkernel}{below=1.1cm of l2m}
\GPSShade{Binnerkernel}{\dmed}{\dbig}{\dmed}{\dbig}{0}{0}{pattern=crosshatch, pattern color=magenta}
\GPSShade{Ainnerkernel}{\dmed}{\dmed}{\dmed}{\dmed}{0}{0}{pattern=crosshatch, pattern color=green}
\GPSPartN{\dmed}{\dbig}{\dmed}{9}{innerkernel}
\GPSBase{\dmed}{\dsma}{\dmed}{nrloop}{below=0.7cm of innerkernel}
\GPSShade{Bnrloop}{\dmed}{\dsma}{\dmed}{\dsma}{0}{0}{pattern=crosshatch, pattern color=cyan}
\GPSShade{Anrloop}{\dmed}{\dmed}{\dmed}{\dmed}{0}{0}{pattern=crosshatch, pattern color=green}
\GPSPartM{\dmed}{\dsma}{\dmed}{3}{nrloop}
\GPSBase{\dsma}{\dsma}{\dmed}{mrloop}{below=0.7cm of nrloop}
\GPSShade{Bmrloop}{\dmed}{\dsma}{\dmed}{\dsma}{0}{0}{pattern=crosshatch, pattern color=cyan}
\GPSShade{Amrloop}{\dsma}{\dmed}{\dsma}{\dmed}{0}{0}{pattern=crosshatch, pattern color=green}
\GPSShade{Cmrloop}{\dsma}{\dsma}{\dsma}{\dsma}{0}{0}{pattern=crosshatch, pattern color=red}
\GPSPartK{\dsma}{\dsma}{\dmed}{12}{mrloop}
%
\path (Cmrloop) +(-3*\dbig/2 + \dmed/2 + \dsma/2, -\dmed/2) coordinate (bottomLeft);
\path (Bl3n) +(\dbig/2 + .3cm, \dbig/2 + .3cm) coordinate (topRightl3n);
\path (Bl3k) +(\dbig/3 + \dbig/3 + .3cm -.04cm, \dbig/2 +.3cm) coordinate (topRightl3k);
\path (Bl2m) +(\dbig/3 + \dbig - \dbig/6 - \dsma/2 + .3cm - .04cm*2, \dbig/2 + .3cm) coordinate (topRightl2m);
\path (Binnerkernel) +(\dbig - \dmed/2 + .3cm - .04cm*3, \dmed/2 + .7cm) coordinate (topRightinnerkernel);
\path (Bnrloop) +(3*\dbig/2 - \dmed/2 - \dsma/2 + .3cm - .04cm*4, \dmed/2 + .3cm) coordinate (topRightnrloop);
\path (Bmrloop) +(3*\dbig/2 - \dmed/2 - \dsma/2 + .3cm - .04cm*5, \dmed/2 + .3cm) coordinate (topRightmrloop);

\draw [rounded corners] (bottomLeft) +(-.3cm, -.3cm) rectangle (topRightl3n);
\draw [rounded corners] (bottomLeft) +(-.3cm + 0.04cm, -.3cm + 0.04cm) rectangle (topRightl3k);
\draw [rounded corners] (bottomLeft) +(-.3cm + 0.04cm*2, -.3cm + 0.04cm*2) rectangle (topRightl2m);
\draw [rounded corners] (bottomLeft) +(-.3cm + 0.04cm*3, -.3cm + 0.04cm*3) rectangle (topRightinnerkernel);
\draw [rounded corners] (bottomLeft) +(-.3cm + 0.04cm*4, -.3cm + 0.04cm*4) rectangle (topRightnrloop);
\draw [rounded corners] (bottomLeft) +(-.3cm + 0.04cm*5, -.3cm + 0.04cm*5) rectangle (topRightmrloop);
%
\path (Cl3n) +(-\dbig/2 - .3cm, \dbig/2 + .3cm) coordinate (topLeftl3n);
\path (Cl3k) +(-\dbig +\dmed/2 - .3cm +.04cm, \dbig/2 +.3cm) coordinate (topLeftl3k);
\path (Cl2m) +(-\dmed/2 - \dbig - .3cm + .04cm*2, \dbig/2 + .3cm) coordinate (topLeftl2m);
\path (Cinnerkernel) +(-\dbig + \dmed/2 - .3cm + .04cm*3, \dmed/2 + .7cm) coordinate (topLeftinnerkernel);
\path (Cnrloop) +(-3*\dbig/2 + \dmed/2 + \dsma/2 - .3cm + .04cm*4, \dmed/2 + .3cm) coordinate (topLeftnrloop);
\path (Cmrloop) +(-3*\dbig/2 + \dmed/2 + \dsma/2 - .3cm + .04cm*5, \dmed/2 + .3cm) coordinate (topLeftmrloop);
\draw [draw=none] (topLeftl3n) -- (topRightl3n) node [midway, above] {\footnotesize{Partition $n$ with blocksize $n_c$}};
\draw [draw=none] (topLeftl3k) -- (topRightl3k) node [midway, above] {\footnotesize{Partition $k$ with blocksize $k_c$}};
\draw [draw=none] (topLeftl2m) -- (topRightl2m) node [midway, above] {\footnotesize{Partition $m$ with blocksize $m_c$}};
\draw [draw=none] (topLeftinnerkernel) -- (topRightinnerkernel) node [midway, above] {\footnotesize{Partition $n$ with blocksize $n_r$}};
\draw [draw=none] (topLeftnrloop) -- (topRightnrloop) node [midway, above] {\footnotesize{Partition $m$ with blocksize $m_r$}};
\draw [draw=none] (topLeftmrloop) -- (topRightmrloop) node [midway, above] {\footnotesize{Micro-kernel}};
%
\path (Bl3k) +(\dbig/3, \dbig/2 - \dsma/2) coordinate (bRight);
\path (Bl2m) +(\dbig/3, 0) coordinate (bTildeRight);
\path (Bl2m) +(\dbig/3 + \dbig/2 - \dsma/2 + .3cm, 0) coordinate (bsdf);
\path (bRight) +(.3cm,0) coordinate (asdf);
\draw [->, line width=1.5pt, darkgray, dashed] (bRight) -- (asdf) -- (bsdf) -- (bTildeRight);
\draw [draw=none] (asdf) -- (bsdf) node [pos=.45, above, rotate=270] {\footnotesize{Pack $\widetilde B$}};
%
\path (Al2m) +(\dsma/2, \dbig/2 - \dsma/2) coordinate (aRight);
\path (Ainnerkernel) +(0, \dmed/2) coordinate (aTildeTop);
\path (aRight) +(3.25cm, 0) coordinate (csdf);
\path (aTildeTop) +(0, +.4cm) coordinate (dsdf);
\path (dsdf) +(3.25cm + \dsma/2,0) coordinate (esdf);
\draw [->, line width=1.5pt, darkgray, dashed] (aRight) -- (csdf) -- (esdf) -- (dsdf) -- (aTildeTop);
\draw [draw=none] (csdf) -- (esdf) node [pos=.60, above, rotate=270] {\footnotesize{Pack $\widetilde A$}};
\end{tikzpicture}
}

\begin{tikzpicture}[>=latex,node distance=-0.25cm and .00cm]
\newcommand{\sWid}{0.4cm}
\node (l3) [pattern=crosshatch, pattern color=magenta,
    fit={(-\sWid/2, -\sWid/2) (\sWid/2, \sWid/2)}, inner sep=0pt, draw=black, thick] {}; 
\node (l2) [pattern=crosshatch, pattern color=green,
    fit={(-\sWid/2, -\sWid/2) (\sWid/2, \sWid/2)}, inner sep=0pt, draw=black, thick, yshift=-0.3cm] at (l3.south) {}; 
\node (l1) [pattern=crosshatch, pattern color=cyan,
    fit={(-\sWid/2, -\sWid/2) (\sWid/2, \sWid/2)}, inner sep=0pt, draw=black, thick, yshift=-0.3cm] at (l2.south) {}; 
\node (l0) [pattern=crosshatch, pattern color=red,
    fit={(-\sWid/2, -\sWid/2) (\sWid/2, \sWid/2)}, inner sep=0pt, draw=black, thick, yshift=-0.3cm] at (l1.south) {}; 
\node [right] at (l3.east) { {\scriptsize Matrix partition is reused in L3 cache.} };  
\node [right] at (l2.east) { {\scriptsize Matrix partition is reused in L2 cache.} };  
\node [right] at (l1.east) { {\scriptsize Matrix partition is reused in L1 cache.} };  
\node [right] at (l0.east) { {\scriptsize Matrix partition is reused in registers.} };  
\end{tikzpicture}

\caption{Diagram of Goto's Algorithm implemented in BLIS.}
\label{fig:goto}
\end{figure}

The algorithm used in the state-of-the-art GotoBLAS~\cite{GotoBLAS} 
and implemented in the BLAS-like Library Instantiation Software (BLIS)~\cite{BLIS1,BLIS2} is illustrated in Figure~\ref{fig:goto}.
Those who are interested in the practical implementation of matrix multiplication should be familiar with these various papers,
and hence we do not go into detail here.
Goto's Algorithm is a practical algorithm that targets multiple layers of cache.
Details of how this algorithm is parameterized based on machine constants can be found in~\cite{BLIS4}.
We will show that the results for this theoretical paper targeting a single layer of cache can 
be used to analyze in what sense Goto's Algorithm does and in what sense it does not achieve optimality with respect to I/O~\footnote{
In our analysis we ignore an extra I/O cost of copying submatrices into contiguous buffers.}.

\paragraph*{Blocking for the L3 cache.}
First, the algorithm partitions the matrices in the $n$ dimension with a blocksize of $n_c$.
Then, it partitions the $k$ dimension with a blocksize of $k_c$, where $k_c $ is determined by further subsequent blocking for the L2 cache.
This bounds the size of a short and very wide panel of $B$ that fill ``most of'' the L3 cache.
Subsequent loops encourage the cache replacement policy to retain that panel of $B$ in the L3 cache.
The number of reads into the L3 cache from main memory is given by 
$\frac{mnk}{n_c} + \frac{mnk}{k_c} + nk$.
This is far from optimal because typically $n_c \gg k_c$ and hence $k_c \ll \sqrt{M_3} $, where $ M_3 $ is the size of the L3 cache.

\paragraph*{Blocking for the L2 cache.}
The algorithm next partitions the $m$ dimension with a blocksize of $m_c$.
This creates a block of $A$ that will reside in the L2 cache, similar to Algorithm A.
If we assume that $n_c$ is very large,
then we can approximate the number of reads into the L2 cache from slower layer of memory by 
$\frac{mnk}{m_c} + \frac{mnk}{k_c} + mk$.
This is close to optimal because typically the block of $A$ in the L2 cache is roughly square and it occupies nearly the entire L2 cache:
In this case $m_c \approx k_c \approx \sqrt{M}$, 
so when $m$, $n$, and $k$ are large we can ignore the $mk$ term and, $\frac{mnk}{m_c} + \frac{mnk}{k_c} \approx \frac{2mnk}{\sqrt{M}}$.

\paragraph*{Subsequent layers of cache.}
For the L1 cache, the algorithm partitions the $n$ dimension with blocksize $n_r$.
This creates a block of $B$ that will reside in the L1 cache, however it is far from square,
and so the number of reads into the L1 cache is suboptimal for similar reasons that the number of reads into the L3 cache is suboptimal.

The final step that we will consider is that algorithm then partitions in the $m$ dimension with blocksize $m_r$.
This creates a roughly square block of $C$ that will reside in registers,
and updated by a series of rank-1 updates, similar to the optimal Algorithm C.

\paragraph*{Summary.}

We conclude that Goto's Algorithm is optimal in the sense that it optimizes for the number of L2 cache misses,
but is suboptimal in terms of L3 cache misses.
Goto's Algorithm is a practical one, and matrix multiplication on modern machines does not need to optimize for L3 cache misses
because there is sufficient bandwidth available from main memory.
However in the future it is possible that even matrix multiplication may become bandwidth limited,
and Goto's Algorithm may need to be tweaked so that it places a square matrix block in the L3 cache.

\subsubsection{A family of algorithms}

Gunnels, et al.~\cite{ITXGEMM} presented a family of algorithms for implementing MMMA.
It discussed three shapes of MMMA that correspond to Algorithms A, B, and C and hence are read-optimal and write-hidden. 
These algorithms result from asking the question of how to optimally block between two adjacent layers of memory and applying this to multiple layers in the hierarchy.  The locally optimal solution is that 
when one of the three algorithms is used for some layer of memory,
one of the other two algorithms should be used at the next faster layer. 
This yields algorithms similar to Algorithms A, B, and C at each level of the cache hierarchy.

\section{Related Work}
We now summarize related work on I/O lower bounds, organizing the works by strategies used in the proofs.

\paragraph{Previous lower bounds on MMM}
Hong and Kung~\cite{redblue} introduced the red-blue pebble game model for a machine with two layers of memory.
A limited number of blue pebbles represented fast memory while an unlimited number of red pebbles represented slow memory.
It is difficult to prove theorems using the red-blue pebble game directly,
so Hong and Kung introduced the S-partitioning problem to reason about I/O complexity instead.
Hong and Kung obtained I/O complexity bounds for several operations. Specifically for MMM, it was of $\Omega({mnk}/{\sqrt{M}})$.
Irony et al.~\cite{irony2004communication} used a simplified version of the S-partitioning problem 
together with the Loomis-Whitney inequality
to obtain a lower bound on the amount of communication between nodes of a distributed memory parallel computer for MMM,
showing that at least one processor must send and receive $(1/2)^{3/2} mnk / (P \sqrt{M})$ elements.
Dongarra et al.~\cite{dongarra2008masterworker} obtained an I/O lower bound for MMM, 
improving the lower bound to $(3/2)^{3/2} mnk / \sqrt{M}$.
Dongarra et al. used a simplified version of the S-partitioning problem and the Loomis-Whitney inequality,
and improved the bound by assuming that all computation is performed via FMA operations.

\paragraph{Other lower bounds from studying S-partitions}
Savage~\cite{savage1995extending} extended the S-partitioning approach to memory hierarchies with more than two levels.
Ballard et al.~\cite{ballard2011minimizing} generalized~\cite{irony2004communication}, showing how the Loomis-Whitney inequality
can be used for essentially any linear algebra operation that can be implemented by a triply-nested loop.
This includes the LU, QR, and Cholesky factorizations, and they obtain lower bounds for each.
The S-partitioning problem and Loomis-Whitney inequality have been applied to find I/O lower bounds for several tensor 
operations~\cite{solomonik2015communication, solomonik2017communication,ballard2017communication}.

Christ, et al.~\cite{christ2013communication} generalized~\cite{ballard2011minimizing}
by using the Holder-Brascamp-Lieb (HBL) inequalities that generalize the Loomis-Whitney inequality.
This provides a methodology for obtaining I/O lower bounds for a wider class of operations than can be reasoned about using the Loomis-Whitney inequality.
Demmel and Dinh~\cite{demmel2018communication} used the Hong-Kung strategy with HBL inequalities to obtain I/O lower bounds
for convolutional neural networks.

\paragraph{Lower bounds from other strategies}
Aggarwal and Vitter~\cite{aggarwal1988input} used a combinatorial argument to find lower bounds for permutation networks,
sorting, matrix transposition, and the fast Fourier transform.
Bilardi, et al.~\cite{bilardi2018lower} use a similar combinatorial argument to
reason about switching DAGs, where the in-degree of a node equals its out-degree.
Bender, et al.\cite{bender2010optimal} used combinatorial arguments to find I/O lower bounds for multiplying a sparse matrix times a dense vector.

Bilardi, et al.~\cite{bilardi1999processor} (which does not allow for recomputation) and Bilardi, et al.~\cite{bilardi2000space} (which does)
introduced a strategy for attaining I/O lower bounds by relating the memory requirements of computations to their I/O requirements.
Elango, et al.~\cite{elango2013data} 
introduced an automated approach to finding I/O lower bounds on a red-blue-white pebble game,
by reasoning about memory requirements of computations.

Aggarwal, et al.~\cite{aggarwal1990communication} introduced the LPRAM model for parallel random access machines with local memory.
They obtain communication lower bounds by reasoning about the critical path length.
Scquizzato, et al.~ \cite{scquizzato2013communication} obtains communication lower bounds for distributed memory computations,
where they do not model fast and slow memory.  Instead they obtain I/O lower bounds by assuming that computation is load-balanced.

\section{Conclusion}
In this paper, we improved the I/O lower bound for $ C := A B + C $ to
${2mnk}/{\sqrt{M}} + mn - 3M$.
We showed that these lower bounds are sharp with respect to the highest ordered term's coefficient 
by analyzing known algorithms.
We also analyzed the state-of-the-art Goto's Algorithm and noted its strengths and weaknesses 
in light of the MMMA lower bound.
These lower bounds are not only of interest as a theoretical result but also to help gain 
fundamental insight into how MMM must be implemented.

We believe that the proof techniques 
presented in this paper can apply to algorithms outside of matrix multiplication.
In particular, in the domain of linear algebra, we believe they can be combined with 
the techniques introduced by Ballard et al.~\cite{ballard2014communication} in order to find lower bounds with improved constants
for other matrix operations such as the LU, QR, and Cholesky factorizations.

\section*{Acknowledgements}
We thank the anonymous referees for their very helpful suggestions.
This work is supported by the National Science Foundation
(grant awards ACI-1550493 and SHF-1645514)
and the Science of High-Performance Computing group's Intel Parallel Computing Center.

\bibliographystyle{ACM-Reference-Format}
\bibliography{biblio}

\appendix

\section{Constrained Global Maximum of $\sqrt{N_A N_B N_C}$}
\label{maximum_xyz}

In this appendix, we give details on how the upper bound $ F_{M+R} $ is determined.
The problem to be solved is
\[
\mbox{maximize }F_{M+R} 
\mbox{ under the constraints }
\left\{ 
\begin{array}{r@{~}c@{~}l}
F_{M+R} &\leq& \sqrt{N_A N_B N_C}\\
0 &\leq& N_A, N_B, N_C \\
N_A + N_B + N_C &\leq& M+R
\end{array} \right. .
\]
If any of $N_A$, $N_B$, or $N_C$ is zero, then so is $F_{M+R}$ and hence will only consider the case where $ 0 < N_A, N_B, N_C $.
If $N_A + N_B + N_C$ are strictly less than $M+R$, then one of $N_A$, $N_B$, or $N_C$ can be increased,
thereby increasing $F_{M+R}$, and hence
we only consider $N_A + N_B + N_C = M+R$.  
Finally, given these constraints we can optimize 
$ F_{M+R} = \sqrt{N_A N_B N_C} $, as long as we check that the result is a maximum.
The constrained problem thus becomes
\[
\mbox{maximize }F_{M+R} = \sqrt{N_A N_B N_C}
\mbox{ under the constraints }
\left\{ 
\begin{array}{r@{~}c@{~}l}
0 & < & N_A, N_B, N_C \\
N_A + N_B + N_C & = & M+R
\end{array} \right. .
\]

\noindent
We can use the Lagrange Multiplier method to solve
$
\nabla F_{M+R} = \lambda \nabla (N_A + N_B + N_C - (M + R)) 
$
for $ N_A $, $ N_B $, $ N_C $.
\[ \frac{N_B N_C}{2\sqrt{N_A N_B N_C}}  = \lambda, \quad
\frac{N_A N_B}{2\sqrt{N_A N_B N_C}}  = \lambda , \quad 
\frac{N_A N_C}{2\sqrt{N_A N_B N_C}}  = \lambda
,\quad \mbox{and }
M + R = N_A + N_B + N_C. \]
Since then $ N_B N_C = N_A N_B = N_A N_C $ and we know that $ N_A $, $ N_B $ and $ N_C $ are nonzero, we deduce that 
$ N_A = N_B = N_C $ and hence 
$ M + R = 3 N_A $.  As a result, the solution is
$ N_A = N_B = N_C = {(M+R)}/{3} $.
To show that this is a global maximum, we can find the second derivative of $F_{M+R}$ at this point,
or we can evaluate $F_{M+R}$ at this point and any point on the boundary of our region to show that any value on the boundary is smaller.

We conclude that the global maximum of $F_{M+R}$ is:
$$ F_{M+R} = \frac{M + R}{3} \sqrt{\frac{M+R}{3}} = \frac{(M+R)\sqrt{M+R}}{3\sqrt{3}}. $$

\end{document}